\documentclass[floats,aps,showpacs,amssymb,
prd,twocolumn,
,secnumarabic%
,tightenlines%
,nofootinbib]{revtex4}%{revtex4-1}
\usepackage{amssymb}
\usepackage{graphics}
\usepackage{amsmath}
\usepackage{amsfonts}
\usepackage{bm}% bold math
\usepackage{graphicx,amssymb,amsmath}
\usepackage{color}

\def\lab{\label}\def\lan{\langle}
\def\lf{\left}
\def\non{\nonumber}
\def\ran{\rangle}\def\rar{\rightarrow}
\def\ri{\right}
\def\al{\alpha}

\def\te{\theta}
\def\si{\sigma}\def\om{\omega}

\begin{document}

\title{Axion--photon mixing in quantum field theory and vacuum energy}

\author{A. Capolupo$^{a,b}$, I. De Martino$^{c,d}$, G. Lambiase$^{a,b}$, and An. Stabile$^{a,b}$}
\affiliation{$^a$Dipartimento di Fisica "E.R. Caianiello" Universit\'a di Salerno}
\affiliation{$^b$INFN - Gruppo Collegato di Salerno, Italy}
\affiliation{$^c$ Donostia International Physics Center (DIPC), 20018 Donostia-San Sebastian (Gipuzkoa)
	Spain}
\affiliation{$^d$ INFN Sez. di Napoli, Compl. Univ. di Monte S. Angelo, Edificio G, Via Cinthia, I-80126, Napoli, Italy}

\date{\today}
\def\be{\begin{equation}}
\def\ee{\end{equation}}
\def\al{\alpha}
\def\bea{\begin{eqnarray}}
\def\eea{\end{eqnarray}}

\begin{abstract}

We analyze axion--photon mixing in the framework of quantum field theory. The condensate structure of the vacuum for mixed fields
induces corrections to the oscillation formulae and leads to non-zero energy of the vacuum for the component of the photon mixed with the axion. This energy generates a new effect   of the vacuum polarization and it has the state equation of the cosmological constant, $w = -1$. This result holds for any homogeneous and isotropic curved space-time, as well as for diagonal metrics.
%The explicit form of the mixed vacuum energy is presented for the Friedmann-Lema\^{i}tre-Robertson-Walker  metric which leads to a time dependent %vacuum energy density.
Numerical estimates of the corrections to the oscillation formulae are presented by considering the intensity of the magnetic field available in the laboratory. Moreover, we estimate the vacuum energy density induced by axion--photon mixing in the Minkowski space-time.
A value compatible with that of the energy density of the universe can be obtained for axions with a mass of   $(10^{-3}-10^{-2}) eV$ in the presence of the strong magnetic fields that characterize astrophysical objects such as pulsars or neutron stars.
In addition, a value of the energy density less than that of the Casimir effect is obtained for magnetic fields used in experiments such as PVLAS. The vacuum polarization induced by this energy  could be detected  in next experiments   and it might provide an indirect proof of the existence of the axion--photon mixing.

The quantum field theory effects presented in this work may lead to new methods for studying axion-like particles.

\end{abstract}


\maketitle

\section{Introduction}

The study of ultra-light particles such as axions, which have very small cross-sections for strong and weak interactions, is a promising field of research that could allow the connection between particle physics and cosmology.
Axions were first proposed to solve the strong CP problem \cite{peccei,peccei1}. They can be described as  boson scalar fields and are expected to have masses in the range of $10^{-6}$ to $10^{-2}$ eV.
Recently, interest has increased in ultra-light axions (ULAs) as candidates for dark matter. ULAs naturally arise in the compactifications of some string landscapes \cite{Witten1984, Hu2000, Svrcek2006, Arvanitaki2010, Marsh2016, Hui2017} and they are expected to have masses of $\sim 10^{-22}$ eV because they are initially massless and acquire it via a symmetry-breaking mechanism \cite{Preskill1983,Abbott1983,Dine1983}.
ULAs and axions have been studied using N-body simulations \cite{Woo2009,Schive2014,Schive2014b,Mocz2017,Veltmaat2018}, interferometry \cite{cap2015}, globular clusters \cite{Hui2016,Emami2018}, galactic pulsars \cite{Khmeinitsky:2013lxt,idm2017,idm2018}, galactic dynamics \cite{idm2018b}, and the EDGES 21-cm absorption signal \cite{bak,lamb}, as well as other probes.

A further property of axions and axion-like particles (ALPs) is axion--photon mixing and oscillation in the presence of strong magnetic fields \cite{raffelt}. This phenomenon is negligible in the case of ULAs. However it could be detected for axions with masses  of $10^{-3}$ and $10^{-2}$ eV.
Many experimental studies have attempted to detect axions \cite{PVLAS1,CAST,CAST1,CAST2,ADMX,ADMX1} and the conversion of photons into axions has also been extensively studied, although without positive results \cite{11,12,13,16,17}.
Despite these efforts, the lack of their direct detection demands the consideration of possible experimental setups that
could demonstrate the existence of axions and ULAs.

In addition to the need to detect the presence of axions, complete theoretical understandings of the behavior of axions and axion--photon mixing are essentials.

In the framework of quantum field theory (QFT), it has been shown that a non-perturbative vacuum structure is associated with the mixing of neutrinos and bosons such as kaons \cite{Blasone:1998hf}-\cite{CapolupoPLB2004}.
 In particular, it was found that the conventional treatment
of neutrino mixing and meson mixing, where the flavor states (i.e., $| \nu_{e}\rangle$, $|\nu_{\mu}\rangle$ and $\nu_{\tau}\rangle$ for neutrinos) are defined in the Fock space of the energy-eigenstates, is affected by the problem of total probability non-conservation, thereby demonstrating that the mixed states should be treated independently from the
energy-eigenstates \cite{Blasone:2005ae}. In fact, the Fock space of the mixed states is unitary inequivalent to the
Fock space of the energy-eigenstates and corrections to the oscillation formulae have been presented.
 The quantum mechanical result is reproduced only in the relativistic limit of
QFT \cite{Blasone:1998hf}-\cite{CapolupoPLB2004}. The contributions to the energy of the universe induced by the condensate structure of the vacuum of mixed particles have also been analyzed \cite{Capolupo:2006etS,Capolupo:2006et2}.

The general theoretical results obtained previously cannot be applied immediately to axion--photon mixing and this phenomenon requires specific adjustments.
Indeed, except for axion--photon mixing, all known mixed systems are characterized by the mixing of particles with the same statistic.
Moreover, vacuum polarization occurs only in the conversion of photons in axions. Therefore, QFT effects on polarization must be considered.

In this paper, we consider axion--photon mixing in the QFT framework. We present new oscillation formulae for the axion--photon system, demonstrate the presence of QFT vacuum polarization induced by the condensate structure of the vacuum for mixed particles, and analyze axion--photon mixing in curved space-time.
We show that the QFT corrections to the amplitudes of the oscillation formulae, which are negligible for neutrinos and kaons, are detectable in principle in the case of axion--photon mixing.
Moreover,
%by considering the Friedmann-Lema\^{i}tre-Robertson-Walker (FLRW) metric,
we show that a non zero vacuum energy with the state equation of the cosmological constant, $w =-1$, characterizes axion--photon mixing.
This energy marks only one component of the photon field and therefore it produces further vacuum polarization.
These results are obtained for a homogeneous and isotropic universe, and for diagonal metrics.
The values of the vacuum energy compatible with that estimated for dark energy can be obtained for axions with masses of $10^{-3}$ to $10^{-2}$ eV in the presence of strong magnetic fields.
In addition, energy density values smaller than that of the Casimir effect might be obtained in the laboratory.
This energy  could be detected  in next experiments by means of the analysis of the vacuum polarization.
 
These contributions are zero in the quantum mechanics mixing treatment. We also verify that all of the QFT effects presented in this study for axions are negligible in the ULA case, where their very low mass inhibits the oscillation with photons.

The  paper is organized as follows. In Section II, we present the QFT treatment of axion--photon mixing and show that the non-perturbative vacuum structure of mixed fields leads to modifications of the standard oscillation formulae and to a new polarization effect of the vacuum.
In Section III, we study axion--photon mixing in curved space. We show that in any homogeneous background and for diagonal metrics, the vacuum energy of the mixed fields assumes a state equation similar to that for the cosmological constant. Moreover, we analyze the energy of the vacuum condensate in flat space-time.
%Axions with mass of order of $(10^{-3}-10^{-2})eV$ in the presence of strong magnetic fields could produce a contribution to the vacuum energy with a value compatible with the one of dark energy of the universe.
We give our conclusions in Section IV.

\section{Axion--photon mixing}

The ALP--photon system is described by the Lagrangian density:
\bea\label{1}
L & = & - \frac{1}{4} F_{\mu \nu} F^{\mu \nu} + \frac{1}{2} (\partial_{\mu} a \partial^{\mu} a - m_{a}^{2} a^{2})
\\\non
& + & \frac{\alpha^{2}}{90 m_{e}^{4}}\lf[(F_{\mu \nu } F^{\mu \nu })^{2}
+ \frac{7}{4}(F_{\mu \nu } \tilde{F}^{\mu \nu })^{2}\ri] + \dfrac{g_{a \gamma \gamma}}{4} \, a \, F_{\mu \nu}\tilde{F}^{\mu \nu}\,,
\eea
where the first two terms are the Lagrangian densities of the free photon and axion, respectively, the third term is the Heisenberg--Euler term due to loop correction in QED \cite{Heisenberg}-\cite{Dobrich},
and the last term is the interaction of two photons with the axion pseudoscalar field $a$ in the presence of a magnetic field \cite{PDG}.
This term is responsible for axion--photon mixing.
Moreover, $\tilde{F}_{\mu \nu } = \frac{1}{2}\epsilon_{\mu \nu \rho \sigma}F^{\rho \sigma} $ is the dual electromagnetic tensor,
$g_{a\gamma \gamma}\equiv g \equiv g_{\gamma} \alpha/ \pi f_{a} $ is the axion--photon coupling with the dimension of the  inverse of the energy, $g_{\gamma} \sim 1$, $\alpha =1/137$ and $f_{a}$ is the decay constant for ALPs. In our treatment, we neglect the Heisenberg--Euler term and
  the birefringence of fluids in a transverse magnetic field (Cotton--Mouton effect).
%can be written as $L_{a \gamma \gamma} = -g_{a \gamma \gamma} a\, {\bf E}  \cdot {\bf B} $.

We consider a monochromatic laser beam  propagating along the z-axes through a region permeated by a magnetic field. We select the y-axis along the projection of $\textbf{B}$ perpendicular to the z-axes (this configuration is typical in laser beams used in   experiments conducted to test the existence of axions \cite{PVLAS1}-\cite{ADMX1}). In this case, the photon polarization state $\gamma_{x} = \gamma_{\bot}$ decouples and the propagation equations can be written as:
\bea
\left(\omega - i \partial_{z} + \textit{M}  \right)\left(
                                                \begin{array}{c}
                                                  \gamma_{\|} \\
                                                  a \\
                                                \end{array}
                                              \right) = 0\,.
\eea
The magnetic field $\textbf{B}$ coincides with the purely transverse field $ {B}_{T}$, and $\textit{M}$ is the mixing matrix:
\bea
\textit{M} = -\frac{1}{2\omega}\, \left(
               \begin{array}{cc}
                 \omega_{P}^{2} & - g \omega B_{T} \\
                 - g \omega B_{T} & m_{a}^{2} \\
               \end{array}
             \right),
\eea
where $\omega_{P} = (4 \pi \alpha N_{e} /m_{e} )^{1/2}$ is the plasma frequency. Here $N_{e}$ is the electron density and $m_{a}$ is the axion mass.
In the case of propagation in the vacuum, we have $\omega_{P} = 0$. However, the transition rate can be enhanced by filling the conversion region with a buffer gas \cite{van} because the presence of the gas induces an effective photon mass given by: $m_{\gamma} =  \frac{\hbar}{c^{2}}\omega_{P}$.
%Therefore, into derive the oscillation formulas, we consider such a case.

 The matrix \textit{M} can be diagonalized by a rotation to  primed fields:
\bea\label{mixing}
\left(
  \begin{array}{c}
    \gamma^{\prime}_{\|}(z) \\
    a^{\prime}(z) \\
  \end{array}
\right) = \left(
            \begin{array}{cc}
              \cos \theta & \sin \theta \\
              -\sin \theta & \cos \theta \\
            \end{array}
          \right)
          \left(
            \begin{array}{c}
              \gamma_{\|}(z) \\
              a(z) \\
            \end{array}
          \right)\,,
\eea
where $\theta = \frac{1}{2} \arctan \displaystyle{\left(\frac{2 g \omega B_{T}}{m_{a}^{2}-\omega^2_P} \right)}$ is the mixing angle,
$\gamma_{\|} $ and $a$ are the fields associated
with the mixed particles, and $\gamma^{\prime}_{\|} $ and $a^{\prime}$ are the ``free''
fields with definite masses ($m_\gamma$ in the presence of plasma, or zero for the photon and $m_a$ for the axion).

 %The evolution of $\gamma^{\prime}_{\|}$ and $a^{\prime}$ is given by
% $\gamma^{\prime}_{\|}(z) = \gamma^{\prime}_{\|}(0) e^{- i \omega_{k,\gamma_{\|}^{\prime}} z}$, $a^{\prime} (z) = a^{\prime} (0) e^{- i \omega_{k,a^{\prime}} z}$,
%with
%$
%\omega_{k, \gamma_{\|}^{\prime}} = \omega_k + \Delta_{-}\,, \quad
%\omega_{k, a^{\prime}} = \omega_k + \Delta_{+}\,
%$, $\omega_k$ photon energy
%and
%\[
%\Delta_{\pm} = - \frac{\omega_{P}^{2}+ m_{a}^{2}}{4 \omega_k} \pm \frac{1}{4 \omega_k} \sqrt{(\omega_{P}^{2}- m_{a}^{2})^{2} + (2 g \omega_k B_{T})^2}\,.
%\]

In order to illustrate the axion--photon mixing in the field-theoretical framework, we invert the mixing relations and consider the following relations:
\bea\non\label{mixingRel}
&&\gamma_{\|} (z) = \gamma_{\|}^{\prime}(z) \; \cos \vartheta + a^{\prime}(z) \; \sin \vartheta
\\[2mm] \lab{2.53}
&& a(z) = - \gamma_{\|}^{\prime}(z) \; \sin\vartheta + a^{\prime}(z)\; \cos\vartheta,
\eea
where $\vartheta = - \theta$.
Notice that any polarization component of a neutral vectorial field  behaves as a neutral scalar field.
Since only the $\gamma_{\|} $ component of the electromagnetic field mixes with the axion field, then
we can treat  $\gamma_{\|} $ as a neutral scalar field.
Therefore, axion--photon mixing can be considered as the mixing of two neutral scalar fields, and thus the formalism presented in Ref.\cite{CapolupoPLB2004} for kaons and the $B_0 - \bar{B}_0 $ system can be applied to the present phenomenon, while considering its intrinsic properties.

We quantize the fields $\gamma^{\prime}_{\|}(x)$ and $a^{\prime}(x)$ in the usual manner, and we recast
Eq. (\ref{2.53}) by means of the generator $G_\vartheta(t)$ (see Appendix 1 for details) in the form:
\bea
\gamma_{\|} (z)  = G^{-1}_\vartheta(t)\; \gamma_{\|}^{\prime} (z)\; G_\vartheta(t) \\[2mm]
\lab{2.53b}
a(z) = G^{-1}_\vartheta(t)\; a^{\prime}(z)\; G_\vartheta(t).
\eea
In the finite volume, $G_\vartheta(t)$ is a unitary operator, which allows us to define
 the vacuum state for mixed fields $\gamma_{||} $ and $a $ as:
$
  |0(\vartheta, t) \ran_{\gamma_{||}, a} \equiv
G^{-1}_ \vartheta(t)\; |0 \ran_{ \gamma_{||}^{\prime},a^{\prime}}\,.
$
This vacuum in the infinite volume limit is orthogonal to $|0  \ran_{\gamma^{\prime}_{||}, a^{\prime}}$, i.e.:
$
\lim\limits_{V\rar \infty}\,_{ \gamma_{||}^{\prime},a^{\prime}}\lan 0|0(\vartheta, t) \ran_{\gamma_{||}, a}
= 0 \,.
$

The annihilation/creation
operators for $\gamma_{\|} $ and $a$ are defined as $\alpha_{{\bf k},\gamma_{\|}}(\vartheta ,t) \equiv G^{-1}_\vartheta(t) \;
\alpha_{{\bf k},\gamma_{\|}^{\prime}}(t)\;G_\vartheta(t)$ and $\alpha_{{\bf k},a}(\vartheta ,t) \equiv G^{-1}_\vartheta(t) \;
\alpha_{{\bf k},a^{\prime}}(t)\;G_\vartheta(t)$,
 such that $\alpha_{{\bf k},\gamma_{\|}}(\vartheta ,t)
|0(\vartheta, t) \ran_{\gamma_{\|}, a} = \alpha_{{\bf k},a}(\vartheta ,t)
|0(\vartheta, t) \ran_{\gamma_{\|}, a} = 0$,
where
$\alpha_{{\bf k},\gamma_{\|}^{\prime}}(t) = \alpha_{{\bf k},\gamma_{\|}^{\prime}}\, e^{-i \om_{k, \gamma_{\|}^{\prime}} t} $,
$ \alpha_{{ \bf k},a^{\prime}}(t) = \alpha _{{ \bf k},a^{\prime}}\, e^{i \om_{k, a^{\prime}} t} $, are the annihilation/creation
operators for the free fields $\gamma_{\|}^{\prime}  $ and $a^{\prime}$, respectively. The energies $\om_{k, \gamma_{\|}^{\prime}}$ and $\om_{k, a^{\prime}}$ are given by:
\bea\label{om1}
\omega_{k, \gamma_{\|}^{\prime}} = \omega_k + \Delta_{+}\,,
\\\label{om2}
\omega_{k, a^{\prime}} = \omega_k + \Delta_{-}\,,
\eea
 with $\omega_k$ photon energy
and
\[
\Delta_{\pm} = - \frac{\omega_{P}^{2}+ m_{a}^{2}}{4 \omega_k} \pm \frac{1}{4 \omega_k} \sqrt{(\omega_{P}^{2}- m_{a}^{2})^{2} + (2 g \omega_k B_{T})^2}\,.
\]
Explicitly, we have:
\bea \label{2.62aneu}\non
\alpha_{{\bf k},\gamma_{\|}}(t)&=&\cos\vartheta  \alpha_{{\bf k},\gamma_{\|}^{\prime}}(t)
\\
&+& \sin\vartheta
\lf( |\Sigma_{{\bf k}}|  \alpha_{{\bf k},a^{\prime}}(t) +
|\Upsilon_{{\bf k}}| \alpha^{\dag}_{{\bf k},a^{\prime}}(t) \ri)\, ,
\\\non
\alpha_{{\bf k},a}(t)&=&\cos\vartheta \alpha_{{\bf k},a^{\prime}}(t)
\\
&-& \sin\vartheta
\lf(|\Sigma_{{\bf k}}| \alpha_{{\bf k},\gamma_{\|}^{\prime}}(t) - |\Upsilon_{{\bf k}}|\;
\alpha^{\dag}_{{\bf k},\gamma_{\|}^{\prime}}(t)\ri)\,,
\eea
where
$ |\Sigma_{{\bf k}}|  $ and $ |\Upsilon_{{\bf k}}|  $
are the Bogoliubov coefficients:
\bea
&&|\Sigma_{{\bf k}}|\equiv \frac{1}{2}
\lf( \sqrt{\frac{\om_{k,\gamma_{||}^{\prime}}}{\om_{k,a^{\prime}}}}
+ \sqrt{\frac{\om_{k,a^{\prime}}}{\om_{k,\gamma_{||}^{\prime}}}}\ri) ~,
\\
&&|\Upsilon_{{\bf k}}|\equiv  \frac{1}{2} \lf(
\sqrt{\frac{\om_{k,\gamma_{||}^{\prime}}}{\om_{k,a^{\prime}}}}
- \sqrt{\frac{\om_{k,a^{\prime}}}{\om_{k,\gamma_{||}^{\prime}}}} \ri)\,,
\eea
which satisfy the following relation:
$|\Sigma_{{\bf k}}|^{2}-|\Upsilon_{{\bf k}}|^{2}=1\,. $

One of the main properties of the vacuum
$|0(\vartheta, t) \ran_{\gamma_{||}, a}$ is its structure of
 condensed particles. For any $t$, the condensation density of $|0(\vartheta, t) \ran_{\gamma_{||}, a}$ is given by:
\bea\label{condensate}
\,_{\gamma_{||}, a}\lan0(\vartheta, t) |\alpha_{{\bf k},\gamma_{||}^{\prime}}^{\dag} \alpha_{{\bf k},\gamma_{||}^{\prime}} |0(\vartheta, t) \ran_{\gamma_{||}, a} = \sin^{2} \vartheta |\Upsilon_{{\bf k}}|^{2}\,,
  \eea
 and a similar result can be obtained by considering $N_{{\bf k},a^{\prime}} = \alpha_{{\bf k},a^{\prime}}^{\dag} \alpha_{{\bf k},a^{\prime}}$.

We note that the condensation density also originates in neutrino and boson mixing cases \cite{Blasone:1998hf}-\cite{CapolupoPLB2004}.
 However, in the present case, the axion only mixes with one of the photon components. This behavior together with photon--axion conversion induces polarization of the   vacuum.

The proposed mechanism can further affect photon vacuum polarization
because the condensate structure of the physical vacuum state defined in Eq. (\ref{condensate})
 only characterizes the parallel component of the photon.
Indeed, the vacuum for axions and photons is given by the following product:
  \bea
  |0(\vartheta,t)\rangle_{\gamma ,a} = |0(\vartheta,t)\rangle_{\gamma_{||},a} \otimes
  |0\rangle_{\gamma^{\prime}_{\bot}},
  \eea
 where $|0(\vartheta,t)\rangle_{\gamma_{||},a}$ is the vacuum for the mixed component of the electromagnetic field $\gamma_{||}$ and $a$, and $|0\rangle_{\gamma^{\prime}_{\bot}}$ is the vacuum for the unmixed component $ {\gamma^{\prime}_{\bot}}$.
 The vacuum $|0\rangle_{\gamma^{\prime}_{\bot}}$ at the difference of $|0(\vartheta,t)\rangle_{\gamma_{||},a}$ has a trivial structure and it does not exhibit a condensed structure. As we show in the following, the vacuum energy is equal to zero
 for $ {\gamma^{\prime}_{\bot}}$. By contrast, $|0(\vartheta,t)\rangle_{\gamma_{||},a}$ lifts the zero point energy to a non-zero and positive value,
   which leads to the production of a foam of condensed particles in the direction parallel to the photon propagation. This condensate is {\it absent} in the direction perpendicular to photon propagation and it derives from the properties of QFT. Indeed, it does not appear in the quantum mechanics mixing treatment.

We now derive the QFT oscillation formulae for the axion--photon system. Assuming an initial state of the photon   $|  \alpha_{{\bf k},\gamma_{\|}}\ran = \alpha^{\dag}_{{\bf k},\gamma_{\|}}(0) |0(\vartheta, 0) \ran_{\gamma_{\|}, a}\,$, the oscillation formulae are obtained by computing the expectation value of the momentum operator for mixed fields ${\cal P}_\si(t)  = \int d^3 x
 \,\pi_{\sigma}(x)\,\nabla\,\phi_{\sigma}(x)$ at $t\neq 0$
normalized to its initial value:
\bea\label{simomenta}
 {{\cal P}}^{\bf k}_{\gamma \rightarrow \si}(t)&\equiv& \frac{  \lan {
\alpha}_{{\bf k},\gamma_{\|}} | {\cal P}_\si(t) | {\alpha}_{{\bf k},\gamma_{\|}}\ran
\,}{  \lan { \alpha}_{{\bf k},\gamma_{\|}} | {\cal P}_\si(0) | {\alpha}_{{\bf
k},\gamma_{\|}}\ran  , }
 \qquad \si = \gamma_{\|}, a.
 \eea
  Explicitly, we have:
  \bea
\label{Amomentum}\non
 {\cal P}^{\bf k}_{\gamma \rightarrow \gamma }(t)&=&
1 - \sin^{2}( 2 \theta) \Big[ |\Sigma_{{\bf k}}|^{2} \; \sin^{2}
\lf( \Omega_{k}^{-} t \ri)
\\
&-& |\Upsilon_{{\bf k}}|^{2} \; \sin^{2} \lf( \Omega_{k}^{+} t
\ri) \Big] \, ,
\\[4mm]\non
{\cal P}^{\bf k}_{ \gamma \rightarrow a}(t)&=&
\sin^{2}( 2 \theta)\Big[ |\Sigma_{{\bf k}}|^{2} \; \sin^{2} \lf(
\Omega_{k}^{-} t \ri)
\\
&-& |\Upsilon_{{\bf k}}|^{2} \;
\sin^{2} \lf( \Omega_{k}^{+} t \ri) \Big]  ,
\label{Bmomentum}
\eea
where $\Omega_{k}^{\pm}=\frac{\omega_{k,a^{\prime}} \pm \omega_{k,\gamma_{\|}^{\prime}}}{2}$.
Eqs. (\ref{Amomentum}) and (\ref{Bmomentum}) are the relative population
densities of the photons and axions in the beam, respectively. These equations indicate the presence of a high-frequency oscillation term that is proportional to:
$|\Upsilon_{\bf k}|^2$.
This term is maximal at low momenta
and tends to zero for large momenta. It depends on the axion mass, coupling constant, and magnetic field.
Moreover, this term is negligible in the case of very low axion masses but in principle, it is detectable in the case of ``heavy'' axions, as shown in Figs (1) and (2). In these figures, we consider laser experiments with a photon energy  $\omega \simeq 1.5 eV$. We assume a coupling constant   $g_{a \gamma} \sim 1.4 \times 10^{-11} GeV^{-1}$ \cite{giannotti} and an axion mass   $m_{a} \in [1 - 10^{-2}] eV$. Notice that a class of axion models allows axion masses up to $0.2 eV$ \cite{iaxo,DiLuzio}.
Figures (1) and (2) show that the presence of plasma reduces the effect of the QFT corrections. In any case, the values obtained for $|\Upsilon_{\bf k}|^2$ are
much higher than the corresponding field-theoretical corrections for the other mixed systems.
\begin{figure}[t]
\begin{picture}(300,180)(0,0)
\put(10,20){\resizebox{8.0 cm}{!}{\includegraphics{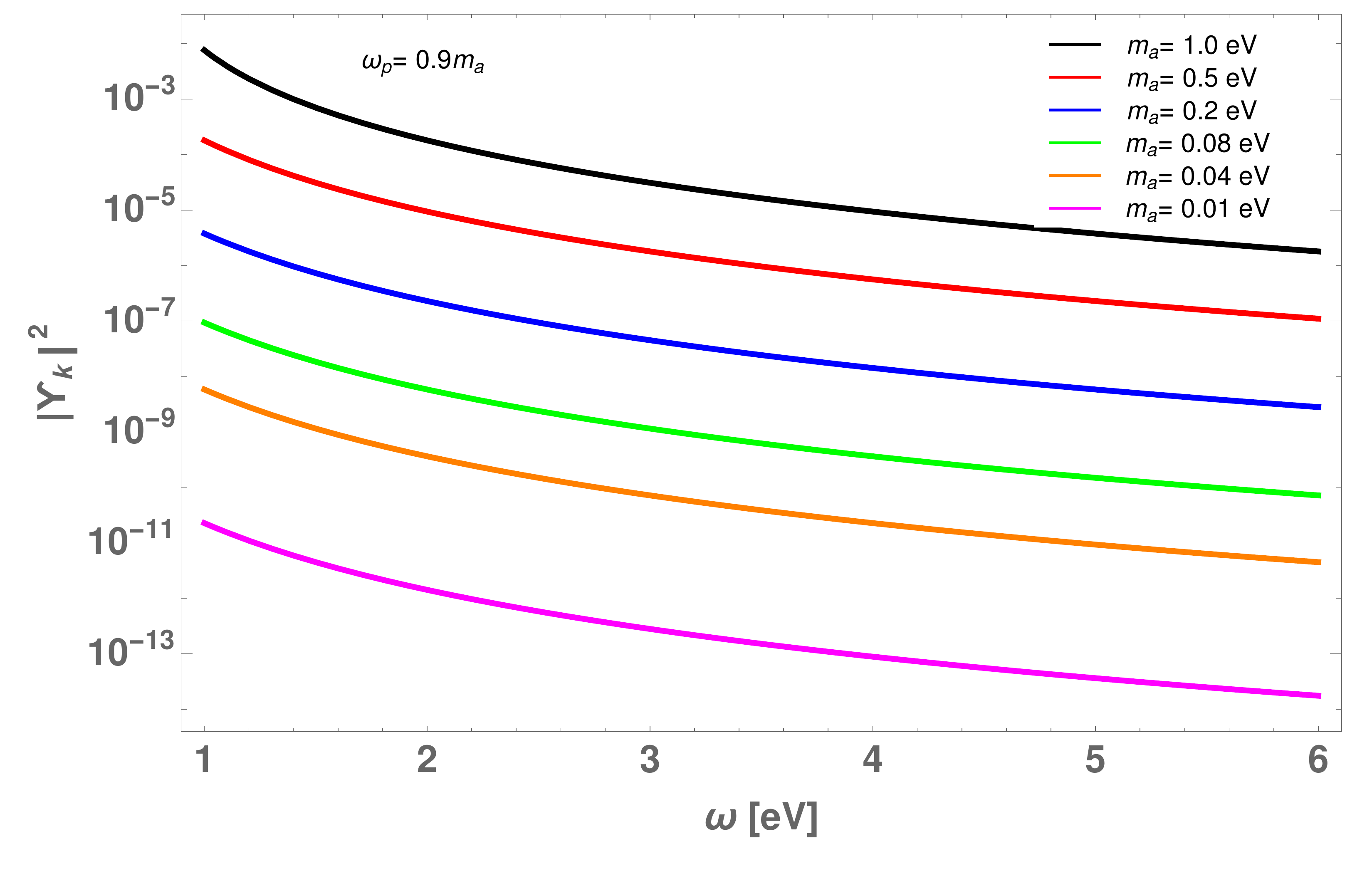}}}
\end{picture}%\vspace{-1cm}
\caption{\em (Color online) Plots of $|\Upsilon_{\bf k}|^2$ as a function of the photon energy for a plasma frequency of $\omega_p = 0.9 m_{a}$.}
\label{pdf}
\end{figure}
\begin{figure}[t]
\begin{picture}(300,180)(0,0)
\put(10,20){\resizebox{8.0 cm}{!}{\includegraphics{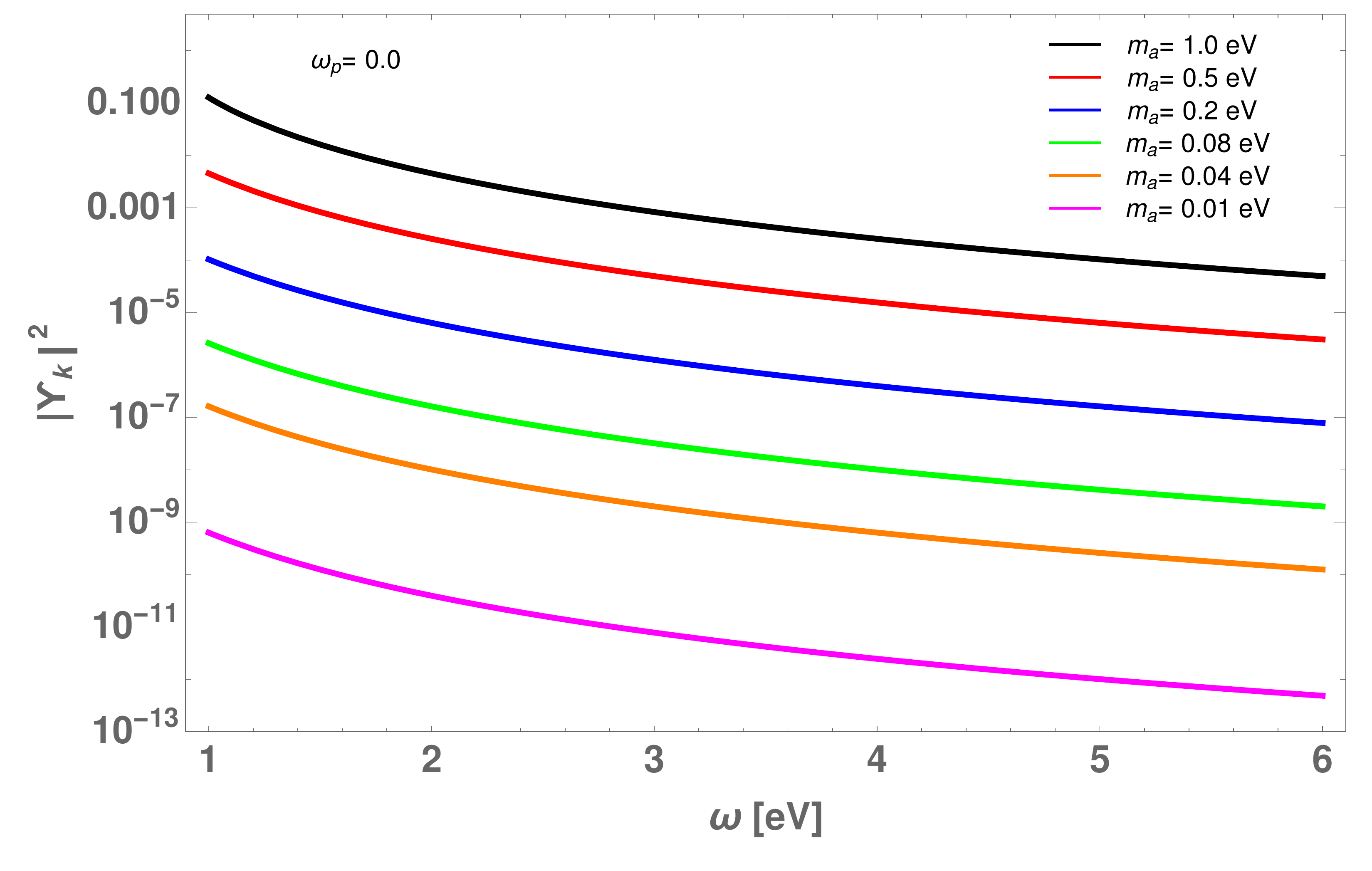}}}
\end{picture}%\vspace{-1cm}
\caption{\em (Color online) Plots of $|\Upsilon_{\bf k}|^2$ as a function of the photon energy for $\omega_p = 0$.}
\label{pdf}
\end{figure}
 Indeed, the values do not exceed
$10^{-26}$ for mixed bosons, such as $K^0 - \bar K^0$, $D^0 - \bar D^0$, $B^0 - \bar B^0$, and $B^0_s -
\bar B^0_s$, and $10^{-18}$ for neutrinos \cite{CapolupoPLB2004}.
Therefore, in principle, the axion--photon system could also represent an interesting system for testing the differences between QFT and quantum mechanics.

The presence of the rapidly oscillating component in Eqs. (\ref{Amomentum}) and (\ref{Bmomentum}) shows that axions and photons behave as two coupled harmonic oscillators.
The origin of this term and the presence of Bogoliubov coefficients in the oscillation formulae are due to the condensate structure of the vacuum for mixed fields. This structure is a genuine field theory phenomenon.
 % fact that when dealing with mixed fields, one intrinsically deals with a manyâ€"particle system, i.e. the mixing is

It should be noted that other rapidly oscillating terms were observed in the quantum field calculation for axion helioscopes   and they correspond  to the emission of ``wrong direction'' photons \cite{Raffelt}. Thus, their nature is completely different from that of the terms presented in this paper.

\section{  Axion--photon mixing in curved space and vacuum energy  }

Now, we analyze the energy of $|0(\vartheta,t)\rangle_{\gamma_{||},a}$ and we show that its state equation has an adiabatic index, $w = -1$.
In particular, we study axion--photon mixing in curved space-time.
In general, cosmic magnetic fields are not uniform on the typical scales of astrophysical objects and $B$ does not have a fixed orientation, but for simplicity, we consider a homogeneous magnetic field which coincides with its transverse component, $B = B_T$.

We quantize the unmixed scalar fields $\psi^{C}_{i} = \gamma^{\prime C}$, $a^{\prime C}$ in curved space-time.
The field operators $\psi_i$ can be expanded in terms of the corresponding
annihilation and creation operators as:
\bea\label{psi}
\psi^{C}_{i}({\bf x},t) = \frac{1}{\sqrt{V}}\sum_{\bf k} \lf[u_{{\bf k},i}({\bf x},t)\alpha_{{\bf k},i} + u^{*}_{{\bf k},i}({\bf x},t)\alpha^{\dag }_{{\bf k},i}  \ri],
\eea
where $u_{{\bf k},i}({\bf x},t) = \zeta_{{\bf k},i}( t) e^{i {\bf k}\cdot{\bf x}}$, and
  the mode functions $\zeta_{{\bf k},i}( t)$ have analytical expressions only in particular cases.
Eq. (\ref{psi}) can be obtained starting from the unmixed scalar fields in flat space-time $\psi_i$, by using the generator $J$ of the curvature:
$\psi^{C}_{i}({\bf x},t) = J^{-1}({\bf x},t)\, \psi_{i}({\bf x},t)\, J({\bf x},t) $. The form of $J$ depends  on the particular metric analyzed \cite{Birrell}. The vacuum state for $\psi^{C}_{i}$ is defined as $|0 ({\bf x},t)\ran^{C}_{ \gamma_{\|}^{ \prime},a^{  \prime}} = J^{-1}({\bf x},t) |0 \ran_{ \gamma_{\|}^{\prime},a^{\prime}}\,.$ Then, the curved mixed vacuum is given by:
%\bea\label{G-curvo}
 $|0(\vartheta, x) \ran^{C}_{\gamma_{\|}, a} =  (G^{C}_ \vartheta(x))^{ -1}\; J^{-1}(x) |0 \ran_{ \gamma_{\|}^{\prime},a^{\prime}}\,,$
%\eea
where $G^{C }_ \vartheta(x)$ is the mixing generator for fields in curved space-time. This generator has a structure similar to the  generator in flat space-time (see Eq. (\ref{2.54neu}) in Appendix 1), but it is expressed in terms of
 the field operators in curved space-time reported in Eq. (\ref{psi}).

For any curved space, the covariant derivative for scalar fields reduces to the ordinary derivative, so the energy momentum tensor density for unmixed scalar fields in curved space time $\psi^{C}_{i}  $ is given by:
\bea
  T_{\mu\nu}(x) & = & \sum_{i = \gamma^{\prime},a^{\prime}}\partial_{\mu}\psi^{C}_{i}(x) \partial_{\nu} \psi^{C}_{i}(x)
 \\ \non
 & - &  \frac{1}{2}  g_{\mu\nu} \Big[\partial^{\rho} \psi^{C}_{i}(x) \partial_{\rho}\psi^{C}_{i}(x)
  - m^{2}(\psi_{i}^{C})^{2}(x)\Big]\,.
  \eea
The contributions to the pressure and energy density are obtained by computing
  the expectation value of $T_{\mu \nu}(x)$ on the curved mixed vacuum:
%\bea\label{G-curvo}
 $|0(\vartheta, x) \ran^{C}_{\gamma_{\|}, a}  ,$
%\eea
i.e.,
\bea\label{exp-val}
\Xi^{\|}_{\mu \nu}  & \equiv & ^{C}_{\gamma_{\|}, a} \langle 0 ( \vartheta, x)|: T _{\mu \nu}(x): | 0 (\vartheta, x)\rangle^{C}_{\gamma_{\|}, a} .
\eea

The symbol, $:...:$, in Eq. (\ref{exp-val}) denotes the normal ordering with respect to the curved vacuum for unmixed fields, $|0 ({\bf x},t)\ran^{C}_{ \gamma_{\|}^{ \prime},a^{  \prime}}$.
In the case of a homogeneous and isotropic universe, as well as for diagonal metrics, the off-diagonal components of $\Xi^{\|} _{\mu \nu}(x)$ are zero \cite{Capolupo:2006etS}, so the condensate behaves as a perfect fluid, and thus we can define the energy density and pressure as:
$
\rho_{\|}  = g^{00}\; \Xi^{\|}_{00} $,
and
$
p_{\|}  = - g^{j j}\; \Xi^{\|}_{ jj } $ (no summation on the index $j$ is intended).
Explicitly, we have (see Appendix 2 for the derivation):
   \bea \label{T00BosMixF}\non
\rho_{\|}   & = &  _{\gamma_{\|}, a} ^{C} \langle 0 ( \vartheta, x)| :\sum_{i=\gamma_{\|}^{\prime},a^{\prime}}
  m^{2}_{i}
(\psi^{C}_{i})^{2}(x) : |0 (\vartheta, x)\rangle^{C}_{\gamma_{\|}, a}\,,
\\
\\\non \label{TjjBosMixF}
p_{\|}  & = & -  \, _{\gamma_{\|}, a} ^{C} \langle 0 ( \vartheta, x)| :\sum_{i=\gamma_{\|}^{\prime},a^{\prime}}
  m^{2}_{i}
(\psi^{C}_{i})^{2}( x) : |0 (\vartheta, x)\rangle^{C}_{\gamma_{\|}, a}\,.
\\ \eea
 Then, the state equation is $w  = - 1$, which is similar to that for the cosmological constant.

We note that nontrivial values of $\rho_{\|}$ and $p_{\|} $ only characterize the component parallel to the magnetic field of the photon polarization because the mixing only involves $\gamma_{\|}$ and $a$.
By contrast, the vacuum energy density and pressure of $\gamma_{\bot}$ are equal to zero:
\bea\label{exp-val2}
\Xi^{\bot}_{\mu \nu}(x) & \equiv & ^{C}_{\gamma_{\bot} } \langle 0({\bf x},t)  |: T _{\mu \nu}(x): | 0({\bf x},t)  \rangle^{C}_{\gamma_{\bot} } = 0 ,
\eea
where $|0 ({\bf x},t)\ran^{C}_{\gamma_{\bot} }= J^{-1}(x)|0 \ran_{\gamma_{\bot} }$ is the vacuum for the component perpendicular to the magnetic field of the photon polarization $\gamma_{\bot}^{C}$, and $|0 \ran_{\gamma_{\bot} }$ is the vacuum for $\gamma_{\bot}$ in the Minkowski metric.
The presence of $\rho_{\|} \neq 0$ for $\gamma_{\|}$ indicates further polarization of the vacuum, which represents a genuine QFT effect.
Indeed, $\rho_{\|}$ is zero in the quantum mechanics axion-photon mixing framework.
If we denote $\Delta m^{2}= |m_{ a^{\prime}}^{2} - m_{\gamma^{\prime}}^{2}|$,
 then we have:
   \bea\label{integral}
 \rho_{\|}  = \frac{\Delta m^{2}   \sin ^{2}\theta}{2} \frac{1}{V}\sum_{{\bf k}}   \lf(|\zeta_{{\bf k},1}( t)|^{2}  - |\zeta_{{\bf k},2}( t)|^{2} \ri),
 \eea
where $V $ is the volume.

It should be noted   that around astrophysical objects such as pulsars or neutron stars where strong magnetic fields are produced (see below), the Schwarzschild metric should be considered:
\bea\label{Sch}
ds^{2} & = & \lf(1 - \frac{2 M}{r} \ri) dt^{2} - \lf(1 - \frac{2 M}{r} \ri)^{-1} dr^{2}
\\ \non
& - & r^{2} (d\theta^{2} + \sin^{2}\theta d\varphi^{2})\,.
\eea
In this case, the Klein--Gordon equation is given in terms of the confluent Heun functions \cite{Bezerra} and the explicit form of the energy density $\rho_{\|}$ is difficult.
In order to derive an estimate of $\rho_{\|}$, we consider the
  Minkowski metric for simplicity, which means that we neglect the gravitational effects induced by the metric in Eq. (\ref{Sch}), i.e., we assume that $r >> 2M$.
   This assumption is reasonable for many astrophysical objects with strong magnetic fields. For example, magnetars have a Schwarzschild radius of about $ 6 km$, and thus the gravitational effect can be neglected for $r >> 6 km $. In the Minkowski metric, the energy density $ \rho_{\|}$ becomes:
   \bea\label{integral2}
 \rho_{\|}  = \frac{\Delta m^{2}   \sin ^{2}\theta}{4 \pi^{2}} \int_{0} ^{K} dk k^{2}  \lf( 1/2\omega_{k,\gamma_{\|}^{\prime}}  - 1/ 2\omega_{k,a^{\prime}}  \ri),
 \eea
 where $K$ is the cut-off on the momenta.
 By considering the expressions for $\omega_{k,\gamma_{\|}^{\prime}}$ and $\omega_{k,a^{\prime}}$ given in Eqs. (\ref{om1}) and (\ref{om2}), and by solving the integral given above analytically, we obtain:
 \bea\label{integral3}\non
 \rho_{\|} & = & \frac{\Delta m^{2}   \sin ^{2}\theta}{8 \pi^{2}} \Big\{  \lf( m_{a^{\prime}}^{2}- \sqrt{ m_{a^{\prime}}^{4}+ 4 g^{2}B_{T}^{2}K^{2}}
 \ri)
 \\
 &+ &\non
 \lf( m_{a^{\prime}}^{2}+ g^{2}B_{T}^{2} \ri)
 \Big[\tanh^{-1} \lf(\frac{\sqrt{ m_{a^{\prime}}^{4}+ 4 g^{2}B_{T}^{2}K^{2}}}{ m_{a^{\prime}}^{2}+ g^{2}B_{T}^{2}} \ri)
\\
 & - &
\tanh^{-1} \lf(\frac{  m_{a^{\prime}}^{2}  }{ m_{a^{\prime}}^{2}+ g^{2}B_{T}^{2}} \ri)  \Big ]\Big\}.
 \eea
For real values of $x$, the function $\tanh^{-1}(x)$ is defined in the domain:
$
-1 < x < 1
$,
(indeed, for $x \in R$, we can write
$ \tanh^{-1}(x) = \frac{1}{2}  \log  \lf( \frac{ 1 + x}{ 1-x} \ri) $), then Eq. (\ref{integral3}) imposes
 a condition on the cut-off $K$ on the momenta given by:
$
-1 < \frac{\sqrt{ m_{a^{\prime}}^{4}+ 4 g^{2}B_{T}^{2}K^{2}}}{ m_{a^{\prime}}^{2}+ g^{2}B_{T}^{2}}< 1 ,
$
which implies the following upper bound on $K$,
\bea\label{K}
K < \frac{\sqrt{2 m_{a^{\prime}}^{2} + g^{2}B_{T}^{2}}}{2}\,.
\eea

We note that the magnetic fields of pulsars and neutron stars are around $10^{12} G$, those of white dwarfs are around $10^6 G$, and
in astrophysical objects, such as active galactic nuclei and quasars,
the magnetic field strength varies in the range of values: $ [10^{6}  -  10^{17}] G$. Other interesting systems include magnetars, which are neutron stars characterized by extremely powerful magnetic fields of $[ 10^{14}-10^{15}]$ G. In these systems, the intense magnetic field powers the emission of electromagnetic radiation, particularly X-rays and gamma rays.

In order to obtain an estimate of $\rho_{\|}$ around the astrophysical systems mentioned above,
%by assuming $\omega_{\gamma^{\prime}} \sim k$ and $\omega_{a^{\prime}} \simeq \sqrt{k^{2} + m^{2}_{a^{\prime}}}$,
we consider an axion mass of $m_{a^{\prime}}\sim 2 \times 10^{-2}eV$, $g \sim 10^{10}GeV^{-1}$, $B_{T} \sim (10^{15} - 10^{16})G $ (which can be obtained for different astrophysical objects), a photon energy of $\omega \sim 100 eV$, and cut-off $K\sim 10^{-2}eV$ (which satisfies Eq. (\ref{K})).
 Thus, we obtain $\rho \sim 10^{-47}GeV^{4}$, which is of the same order as the estimated value of the dark energy.
This contribution to the vacuum energy is produced around astrophysical objects with high magnetic field. The highly magnetized volume around these objects is very small compared with intergalactic scales. Moreover, the energy density is not made much larger than that of dark energy, so this contribution is negligible around such astrophysical systems. In addition, these objects are generally bound inside galaxies where dark energy and vacuum energy are generally not observable. We also note that the plasma frequencies, photon energies $\omega$, and the mixing angle $\theta_{a}$ depend on the particular system considered. In general, small fluctuations from the parameter values lead to different values for $\rho $.
%However, the energy of the vacuum condensate produced in such systems could represents a "new" contribution to the total vacuum energy of the universe.

Let us now consider terrestrial experiments. Strongest continuous magnetic field yet produced in a laboratory is $45 T$. Due to the presence of the term  $\tanh^{-1} \lf(\frac{  m_{a^{\prime}}^{2}  }{ m_{a^{\prime}}^{2}+ g^{2}B_{T}^{2}} \ri)$ in Eq. (\ref{integral3}), for $g \sim 10^{10}GeV^{-1}$, the energy density $\rho_{\|}$ has a finite value only for axion mass values less than $3 \times 10^{-7}eV.$
For example, for $m_{a^{\prime}}\sim 2 \times 10^{-7}eV$ and $\omega \sim 10 eV$, we have $\rho \sim 10^{-66}GeV^{4}$. This energy density is less than  the Casimir  energy  density. Indeed, for a length $L$ between parallel plates   of about $ (10^{-5} -  10^{-4}) m$, the energy density generated by the Casimir effect is $E=\frac{\pi^2}{1440 L^4} =   (10^{-44} - 10^{-48})GeV^{4}$, respectively.
However, the vacuum polarization induced by $\rho_{\|} $  could be detected  in experiments similar to PVLAS   \cite{PVLAS}  and might represent an indirect proof of the existence of the axion--photon mixing.

It should be noted that the new effects revealed in this work by analyzing axion--photon mixing in the QFT framework (i.e., corrections to the oscillation formulae and a non-trivial vacuum energy that produces further vacuum polarization) are not expected in the usual quantum mechanics treatment of axion--photon mixing.

\section{Conclusions}

We considered many aspects of the QFT formalism for
axion--photon mixing.
The unitary inequivalence between
the space for the mixed field states and the state space where the unmixed field operators are
defined, affects the oscillation formulae and leads to polarization of the vacuum.
These effects, that are not expected in previous studies of axion--photon mixing in the framework of quantum mechanics,
in principle, they can be detected in laser beam experiments.
In order to connect our findings with large-scale structures such as cosmology or astrophysics frameworks,
 we considered axion--photon mixing in curved space-time.
  We showed that the vacuum energy induced by axion-photon mixing for a homogeneous and isotropic universe as well as for diagonal metrics has a state equation with the adiabatic index $w = -1$.
We provided a numerical estimate of the energy density for the vacuum condensate for the Minkowski metric. A value compatible with that for dark energy can be obtained for axions with masses of $10^{-3} - 10^{-2} eV$, in regions of the universe with strong magnetic fields. However, this energy does not contribute to the cosmological dark energy, and it is a purely local phenomenon that only appears in the magnetized volume around astrophysical objects where other fields would have a larger impact on the energy-momentum tensor. For example, the energy density of magnetic fields $B$ with values $   (10^{15} - 10^{16}) G$ is $\rho_{B} \in (10^{-10} - 10^{-8}) GeV^{4}$.

On the other hand, in principle, the energy density of the vacuum condensate induced by axion--photon mixing and QFT vacuum polarization can be detected in laser beam experiments for ALPs with masses less than $10^{-7}eV$.

The theoretical results presented in this paper could open new interesting scenarios in the research
into axions and axion-like particles.

%%%%%%%%%%%%%%%%%%%%%%%%%%%%%%%
\section*{Acknowledgments}
%%%%%%%%%%%%%%%%%%%%%%%%%%%%%%%%%%

AC, GL, and AS acknowledge partial financial support from MIUR and INFN.
AC, IDM and  GL acknowledge   the COST Action CA1511 Cosmology
and Astrophysics Network for Theoretical Advances and Training Actions (CANTATA)
supported by COST (European Cooperation in Science and Technology).

%\vspace{0.2in}

\section*{Appendix A: Axion--photon mixing in the Minkowski metric - useful formulae}

The Fourier expansions of the free fields $\gamma^{\prime}_{\|} $ and their
conjugate momenta $\pi^{\prime}_{\gamma}$ in flat space-time are:
\bea\lab{2.51} \non
\!\gamma^{\prime}_{\|}(x) = \int \frac{d^3k}{(2\pi)^{\frac{3}{2}}}
\frac{1}{\sqrt{2\om_{k,\gamma_{\|}^{\prime}}}} \lf( \alpha_{{\bf k},\gamma_{\|}^{\prime}}(t) + \alpha^{\dag }_{{-\bf k},\gamma_{\|}^{\prime}}(t)  \ri)
e^{i {\bf k}\cdot {\bf x}}
\\
\\\non\lab{2.52}
\pi^{\prime}_{\gamma}(x) = i\,\int \frac{d^3 k}{(2\pi)^{\frac{3}{2}}}
\sqrt{\frac{\om_{k,\gamma_{\|}^{\prime}}}{2}} \lf( \alpha^{\dag}_{{\bf k},\gamma_{\|}^{\prime}}(t)
  - \alpha_{{-\bf k},\gamma_{\|}^{\prime}}(t) \ri)
e^{i {\bf k}\cdot {\bf x}}.
\\
\eea
Similar expansions hold for the free axion fields $a^{\prime} (x)$
and their conjugate momenta $\pi^{\prime}_{a}$.
In Eqs. (\ref{2.51}) and (\ref{2.52}), we have $\alpha_{{\bf k},\gamma_{\|}^{\prime}}(t) = \alpha_{{\bf k},\gamma_{\|}^{\prime}}(0)e^{- i \omega_{k,\gamma_{\|}^{\prime}} t}$, and similarly, the time evolution of the annihilators for axions is:
$\alpha_{{\bf k},a^{\prime}}(t) = \alpha_{{\bf k},a^{\prime}}(0)e^{- i \omega_{k,a^{\prime}} t}$.

The form of the generator $G_\vartheta(t)$ of the mixing transformation is:
\bea\lab{2.54neu}
G_\vartheta(t) = exp\lf[-i\;\vartheta \int d^{3}x
\lf(\pi^{\prime}_{\gamma}(x) a^{\prime}(x) - \gamma_{\|}^{\prime}(x)\pi^{\prime}_{a}(x)\ri)\ri]\, ,
\eea
which can be written as:
\bea\lab{2.55neu} \non
G_\vartheta(t) & = & \exp \Big[\vartheta   \int d^3 k \Big( |\Sigma_{{\bf k}}|\,
\alpha_{{\bf k},\gamma_{||}^{\prime}}^{\dag}(t) \alpha_{{\bf k},a^{\prime}}(t)
\\\non
 &-& |\Upsilon_{{\bf k}}|   \,
\alpha_{{\bf k},\gamma_{||}^{\prime}}(t) \alpha_{{\bf k},a^{\prime}}(t)
  +   |\Upsilon_{{\bf k}}|  \,
\alpha_{{\bf k},\gamma_{||}^{\prime}}^{\dag}(t) \alpha_{{\bf k},a^{\prime}}^{\dag}(t)
 \\
 &-& |\Sigma_{{\bf k}} | \,
\alpha_{{\bf k},\gamma_{||}^{\prime}}(t) \alpha_{{\bf k},a^{\prime}}^{\dag}(t)
\Big)\Big].
\eea

 The orthogonality between $|0(\vartheta, t) \ran_{\gamma_{||}, a}$ and $| 0\ran_{ \gamma_{||}^{\prime},a^{\prime}}$ in the infinite volume limit can be shown easily by considering the explicit form of the product $\,_{ \gamma_{||}^{\prime},a^{\prime}}\lan 0|0(\vartheta, t) \ran_{\gamma_{||}, a}$.
Indeed, we have $ \lim\limits_{V\rar \infty}\,_{ \gamma_{||}^{\prime},a^{\prime}}\lan 0|0(\vartheta, t) \ran_{\gamma_{||}, a}
= \lim\limits_{V\rar \infty}\, e^{\frac{V}{(2\pi)^3}
\int d^3 k \, \ln
\lf(\frac{1}{1 + |\Upsilon_{\bf k}|^{2} \sin^{2}\vartheta} \ri) } \, = \, 0 ~, $ since the logarithm is negative for any values of ${\bf k}$, $\te$ and
$m_\gamma, m_a$ (note that $0\le \vartheta \le \pi/4$).

The explicit form of the momentum operator for mixed fields is:
\bea
{{\cal P}}_{\sigma}(t) = \int d^3 k \frac{{{\bf
k}}}{2} \Big(\alpha^{\dag}_{{\bf k},\sigma}(t) \alpha_{{\bf k},\sigma}(t)-
\alpha^{\dag}_{{-\bf k},\sigma}(t) \alpha_{{-\bf k},\sigma}(t)\Big),
\eea
where $ {\sigma} = \gamma_{\|}, a$.

\section*{Appendix B: Axion--photon mixing in curved space - derivation of $\rho_{\|}$ and $p_{\|}$ }

For a homogeneous and isotropic universe as well as for diagonal metrics, the energy density $\rho_{\|}$ and pressure $p_{\|}$ of the vacuum condensate induced by axion--photon mixing are given by:
\begin{widetext}

\bea \label{T00BosGen}\non
\rho_{\|}  & = &   g^{00}  \, _{\gamma_{\|}, a} ^{C} \langle 0 ( \vartheta, t)| : \sum_{i=\gamma_{\|}^{\prime},a^{\prime}} \Big[   \lf[\partial_{0}\psi_{i}^{C}( x)\ri]^{2}    - \frac{1}{2}  g_{00} \Big(g^{00}\lf[\partial_{0}\psi_{i}^{C}( x)\ri]^{2}
+
 g^{kk}\partial_{k}\psi_{i}^{C}( x) \partial_{k}\psi_{i}^{C}( x)
- (m_{i} \psi_{i}^{C})^{2}( x) \Big)\Big]: | 0 (\vartheta, t)\rangle^{C}_{\gamma_{\|}, a}\,;
\\
\\
\label{TjjBosGen}\non
p_{\|}  & = & -g^{jj}\, _{\gamma_{\|}, a} ^{C} \langle 0 ( \vartheta, t)| :\sum_{i=\gamma_{\|}^{\prime},a^{\prime}}\Big[ \lf[ \partial_{j} \psi_{i}^{C}( x) \ri]^{2}- \frac{1}{2}  g_{jj} \Big(g^{00}\lf[\partial_{0}\psi_{i}^{C}( x)\ri]^{2}
+
 g^{kk}\partial_{k}\psi_{i}^{C}( x) \partial_{k}\psi_{i}^{C}( x)
 - (m_{i}
\psi_{i}^{C})^{2}( x) \Big) \Big]: | 0 (\vartheta, t)\rangle^{C}_{\gamma_{\|}, a}\,.
\\ \eea
\end{widetext}

Let us denote with $ \psi_\sigma(\vartheta,x)$ the mixed fields, where $\sigma = \gamma_{\|},a$. For any homogeneous and isotropic metric, by considering the mixing relations in Eq. (\ref{mixingRel}), at any cosmic time, we have:
\bea\label{a1}
\sum_{\sigma=\gamma_{\|},a}   \psi^{2}_\sigma(\vartheta,x) & = & \sum_{i=\gamma_{\|}^{\prime},a^{\prime}}  \psi^{2}_{i}(x)\,,
\\\label{a2}
\sum_{\sigma=\gamma_{\|},a} [\partial_{0}\psi_{\sigma}(\vartheta,x)]^{2} & = & \sum_{i=\gamma_{\|}^{\prime},a^{\prime}}  [\partial_{0} \psi_{i}(x)]^{2}\,,
\\\label{a3}
\sum_{\sigma=\gamma_{\|},a} [\partial_{j}\psi_{\sigma}(\vartheta,x)]^{2} & = & \sum_{i=\gamma_{\|}^{\prime},a^{\prime}}  [\partial_{j} \psi_{i}(x)]^{2}\,,
   \eea
  i.e., these operators are invariant under the action of the generator $G ^{C}_ \vartheta(t)$,
  which implies that the kinetic and gradient terms of the mixed vacuum are equal to zero:
  \bea \label{T00Bosvac}\non
      \, _{\gamma_{\|}, a} ^{C} \langle 0 ( \vartheta, t)| : \sum_{i=\gamma_{\|}^{\prime},a^{\prime}}     \lf[\partial_{0}\psi_{i}( x)\ri]^{2} : | 0 (\vartheta, t)\rangle^{C}_{\gamma_{\|}, a} &=& 0
  \\\non
  \, _{\gamma_{\|}, a} ^{C} \langle 0 ( \vartheta, t)| : \sum_{i=\gamma_{\|}^{\prime},a^{\prime}}
  \lf[\partial_{k}\phi_{i}( x)\ri]^{2} : | 0 (\vartheta, t)\rangle^{C}_{\gamma_{\|}, a} &=& 0 .
   \eea
   On the contrary, the mass terms in Eqs. (\ref{T00BosGen}) and (\ref{TjjBosGen}) generate non-vanishing contributions
   because $m_{\gamma} \neq m_{a}$. From the relations  given above, we obtain Eqs. (\ref{T00BosMixF}) and (\ref{TjjBosMixF}) for
   $\rho_{\|}$ and $p_{\|}$.

\end{document}